\documentclass{moriond}

\usepackage{amsmath,amssymb}
\def\be{\begin{equation}}
\def\ee{\end{equation}}
\def\bea{\begin{eqnarray}}
\def\eea{\end{eqnarray}}

\newcommand{\Photo}{}

\begin{document}
\title{Modern techniques of multiloop calculations}

\author{ R.N. Lee }

\address{The Budker Institute of Nuclear Physics, 630090 Novosibirsk, Russia}

\maketitle\abstracts{I present a few new and recent ideas of the multiloop calculations.}

\section{Introduction}

New Physics is expected to show up as tiny deviations from the predictions of the Standard Model. Therefore, its searches essentially depend on our ability to obtain high-precision theoretical predictions within the Standard Model, and, in general, in Quantum Field Theory. This is where the multiloop calculations come into play. The complexity of these calculations rapidly grows with the increase of the the number of loop and/or legs. Therefore, there is a constant demand for the new methods of the multiloop calculations. Another important issue is the necessity of the automatization of as many stages of the calculation as possible. 


As it concerns the automatization, one of the most fruitful idea is the use of the integration-by-part (IBP) identities \cite{Chetyrkin1981,Tkachov1981} in order to reduce the problem to the calculation of a finite set of the master integrals. While the derivation of the IBP identities is simple, their effective use for the reduction is still an open problem.  Despite more than thirty years of using the IBP reduction, there is no known universal algorithm of finding all necessary reduction rules. Practical approaches implemented in most public packages mainly rely on the brute-force search, known as Laporta algorithm \cite{Laporta2000}.

It is important that the IBP reduction allows one to obtain the differential \cite{Kotikov1991,Remiddi1997} and difference~\cite{Laporta2000,Tarasov1996,Lee2010} equations for the master integrals. These equations can be used for the calculation of the master integrals without explicit integration. The method of differential equations can not be directly applied to the integrals depending on one scale. In this case one may switch to the difference equations. In particular, in dimensional regularization there is an elegant possibility to use the dimensional shift relations \cite{DerkachovHonkonenPismak1990,Tarasov1996}.

\section{Multiloop integral}\label{sec:ml-integral}
To fix notation, let us assume that we are interested in the contribution of the $L$-loop diagram depending on the $E$ external momenta $p_{1},\ldots,p_{E}$. There are $N=L(L+1)/2+LE$ scalar products depending on the loop momenta $l_{i}$:
\begin{equation}
s_{ij}=l_{i}\cdot q_{j}\,,\ 1\leqslant i\leqslant L,j\leqslant L+E,
\end{equation}
where $q_{1,\ldots,L}=l_{1,\ldots,L}$, $q_{L+1,\ldots,L+E}=p_{1,\ldots,E}$. After the tensor reduction and partial fractioning, the problem is reduced to the calculation of the scalar integrals of the following form
\begin{align}
J^{(d)}\left(\boldsymbol{n}\right)&=J(n_{1},n_{2},\ldots,n_{N})=(i\pi^{d/2})^{-L}\int\frac{ d^{d}l_{1}\ldots d^{d}l_{L}}{D_{1}^{n_{1}}D_{2}^{n_{2}}\ldots D_{N}^{n_{N}}}\,,\label{eq:J}
\end{align}
As usual, we assume that $D_1,\ldots,D_N$ form a complete basis in the sense that any $s_{ik}$ can be uniquely expressed in terms of $D_{\alpha}$. Some of $D_{\alpha}$ correspond to the denominators of the propagators, others may correspond to the irreducible numerators. E.g., the $K$-legged $L$-loop diagram with generic external momenta corresponds to $E=K-1$ and the maximal number of denominators is $M=E+3L-2$, so that the rest $N-M=(L-1)(L+2E-4)/2$ functions correspond to irreducible numerators.   The integral $J^{(d)}\left(\boldsymbol{n}\right)$ is said to belong to the {\em sector} $\boldsymbol{\theta}=(\theta_1,\ldots,\theta_N)$, where $\theta_\alpha=\theta(n_\alpha-1/2)$. The multiindex $\boldsymbol{n}=(n_1,\ldots,n_N)$ can be thought of as a point in $\mathbb{Z}^{N}$ and the whole family \eqref{eq:J} determines a function $J^{(d)}: \mathbb{Z}^{N}\to\mathbb{C}$. In what follows, we will use operators $A_{\alpha}$ and $B_{\alpha}$ acting on such functions and defined as
\begin{align}\label{eq:AB}
\left(A_{\alpha}J^{(d)}\right)\left(n_{1},\ldots,n_{N}\right) & =n_{\alpha}J^{(d)}\left(n_{1},\ldots,n_{\alpha}+1,\ldots,n_{N}\right),\nonumber \\
\left(B_{\alpha}J^{(d)}\right)\left(n_{1},\ldots,n_{N}\right) & =J^{(d)}\left(n_{1},\ldots,n_{\alpha}-1,\ldots,n_{N}\right).
\end{align}
\section{Reduction and parametric representation}


The conventional form of the parametric representation of the integral $J\left(\mathbf{n}\right)$ depends on two polynomials
\begin{equation}
U(z_1,\ldots,z_N)=\det\left(a\right),\quad F(z_1,\ldots,z_N)=c\det\left(a\right) -\left(a^{\mathrm{Adj}}\right)^{ij}b^{i}\cdot b^{j},
\end{equation}
where $a^{ij}=a^{ji}$, $b^{i}$, and $c$ are determined from
\begin{equation}
\sum_{\alpha}z_{\alpha}D_{\alpha}=\sum_{i,j}a^{ij} l_i\cdot l_j+2 \sum_i b^{i}\cdot l_i+c\,.
\end{equation}
In Ref. \cite{LeePomeransky2013} the modified parametric representation has been introduced. The resulting formula, which works also for the integrals with numerators, reads
\begin{equation}\label{eq:FPG}
J^{(d)}(\mathbf{n})=
\frac{\Gamma\left[d/2\right]}{\Gamma\left[\left(L+1\right)d/2-\Sigma n\right]}\prod_{\alpha}I_\alpha^{n_\alpha} G^{-d/2}\,,
\quad G=F+U\,.
\end{equation}
where $\Sigma n=n_1+\ldots n_N$, and the functionals $I_\alpha^{m}$ are determined as 
\begin{equation}
I_\alpha^{m}[\phi(z_\alpha)]=\left\{\begin{array}{rl}
\int_0^{\infty} \frac{dz_\alpha z_\alpha^{m-1}}{\Gamma(m)}\phi(z_\alpha) & m>0\\
(-1)^{m}\phi^{(-m)}(0) & m\leqslant 0
\end{array}\right.
\end{equation}
for smooth functions falling off sufficiently fast at $z_\alpha \to\infty$.
Note that these functionals satisfy relations
\begin{equation}\label{eq:Iproperties}
I_\alpha^{m}[-\partial\phi(z_\alpha)/\partial z_\alpha]=I_\alpha^{m-1}[\phi(z_\alpha)]\,,\quad
I_\alpha^{m}[z_\alpha\phi(z_\alpha)]=m I_\alpha^{m+1}[\phi(z_\alpha)]\,.
\end{equation}

In Ref. \cite{LeePomeransky2013} it was shown that the number of master integrals in a given sector $\boldsymbol{\theta}$
can be determined from the properties of the polynomial $G_{\boldsymbol{\theta}}$, where $G_{\boldsymbol{\theta}}$ is obtained from $G$ by nullifying all parameters corresponding to the numerators of this sector. Loosely speaking, the main idea of the consideration of Ref. \cite{LeePomeransky2013} was the following. The IBP identities may be thought of as defining equivalences between exterior differential forms up to total derivative. The master integrals are then naturally associated with the basis of the de Rham cohomology group. Their number is just the vector space dimensionality of this group (the Betti number). Due to the Poincare duality, this dimensionality should be equal to the number of independent integration cycles. Strictly speaking, this duality necessarily holds only for compact manifolds, so some complications connected with non-compactness of $\mathbb{C}^M$ are expected. Apart from these complications, the basis of the integration cycles can be chosen as the upward flow contours attached to the nonzero critical points of  $G_{\boldsymbol{\theta}}$. The number of these critical points can be calculated algebraically, and the {\em Mathematica} package \texttt{Mint} has been derived to automatize this task.

Note that the above consideration tacitly assumes the possibility to derive IBP relations directly in the parametric representation. Besides, it only gives one a tool to count the master integrals from the parametric representation, but not to make the reduction. Therefore, the explicit form of the IBP relations in the parametric representation is of a considerable interest. 

In order to derive these relations, one might try the following approach. Start from the expression 
\begin{equation}\label{eq:expr}
I_1^{n_1}\ldots I_N^{n_N}[-\partial_\alpha (G/G^{d/2})]\,.
\end{equation} 
Due to the first relation \eqref{eq:Iproperties}, this expression is equal to $
I_1^{n_1}\ldots I_\alpha^{n_\alpha+1}\ldots I_N^{n_N}[G/ G^{d/2}]$. On the other hand, explicit differentiation gives
$(d/2-1)I_1^{n_1}\ldots I_N^{n_N}[(\partial_\alpha G)/ G^{d/2}]$. Using second relation \eqref{eq:Iproperties}, it is easy to express both forms via integrals $J^{(d)}$ with shifted indices. However, it can be checked explicitly that the obtained identities do not give satisfactory reduction. It means that putting $G$ in the numerator of Eq.\eqref{eq:expr} was too generous, and we could be more economic. Indeed, assume that polynomials $Q(z_1,\ldots,z_N),\,Q_1(z_1,\ldots,z_N),\,\ldots Q_N(z_1,\ldots,z_N)$ are such that the following relation holds
\begin{equation}\label{eq:syzygyrel}
Q G + \sum_{\alpha} Q_\alpha \partial_\alpha G=0\,.
\end{equation}
Then, starting from $I_1^{n_1}\ldots I_N^{n_N}[-\sum_{\alpha}\partial_\alpha (Q_\alpha/G^{d/2})]$ instead of \eqref{eq:expr}, we still obtain relations between $J^{(d)}$. The resulting identity can be represented as
\begin{equation}\label{eq:ibp}
\left(\sum_{\alpha}  Q_\alpha(A_1,\ldots,A_N)B_\alpha+\frac{d}2Q(A_1,\ldots,A_N)]\tilde{J}^{(d)}\right)(\mathbf{n})=0\,,
\end{equation}
where $A_\alpha$ and $B_\alpha$ are defined in Eq. \eqref{eq:AB}, and $\tilde{J}^{(d)}(\mathbf{n})=\Gamma\left[\left(L+1\right)d/2-\Sigma n\right]J^{(d)}(\mathbf{n})$.
Note that \eqref{eq:expr} corresponds to a specific choice $Q=-\partial_\alpha G,\,Q_\beta=G \delta_{\alpha\beta}$.

The $(N+1)$-tuple $\langle Q,\,Q_1,\,\ldots Q_N \rangle$ satisfying Eq. \eqref{eq:syzygyrel} is called a {\em syzygy} of the module $M$ generated by  $\langle G,\,\partial_1 G,\,\ldots \partial_N G\rangle$. The set of all syzygies form a {\em syzygy module} of $M$. It is important that, given $M$, one can routinely find the Groebner basis of the syzygy module. The corresponding algorithms are implemented in many computer algebra systems. For each element of the basis we can construct the identity \eqref{eq:ibp} and this will give us the IBP identities in the parametric representation. Some tests undertaken show good perspectives of the reduction based on the described approach. The details of this approach will be described elsewhere.

\section{Differential equations}

The general idea of constructing the differential equations is to differentiate the master integral with respect to some external parameter under the integral sign and to make the IBP reduction. Differential equations with respect to the invariant constructed of the external momenta \cite{Remiddi1997,Gehrmann2000} can be obtained using the following formulas:
\begin{align}
\frac{\partial}{\partial\left(p_{1}\cdot p_{2}\right)}J\left(\mathbf{n}\right) & =\sum\left[\mathbb{G}^{-1}\right]_{i2}p_{i}\cdot\partial_{p_{1}}J\left(\mathbf{n}\right),\quad 
\frac{\partial}{\partial\left(p_{1}^{2}\right)}J\left(\mathbf{n}\right) & =\frac{1}{2}\sum\left[\mathbb{G}^{-1}\right]_{i1}p_{i}\cdot\partial_{p_{1}}J\left(\mathbf{n}\right),\label{eq:Dinv}
\end{align}
where $\mathbb{G}=\{p_{i}\cdot p_j\}$ is a Gram matrix. After the IBP reduction the right-hand side is expressed in terms of the master integrals. In the general case of several parameters $x,y,\ldots$ the resulting differential system can written in the form
\begin{equation}
d \mathbf{J} =\mathbb{M}(\epsilon,x,y,\ldots) \mathbf{J}\,,
\end{equation}
where $\mathbf{J}$ is the column of the master integrals and $\mathbb{M}=\mathbb{M}_x dx+\mathbb{M}_y dy+\ldots$ is some differential form with matrix coefficients.

In Ref. \cite{Henn2013} a remarkable observation has been made: with a wise choice of the master integrals the $\epsilon$-dependence of $\mathbb{M}$ is reduced to the overall factor $\epsilon$. This factorization dramatically simplifies solution of the differential system which was demonstrated in many applications \cite{HennSmirnov2013,HennSmirnovSmirnov2014,HennMelnikovSmirnov2014,CaolaHennMelnikovSmirnov2014,GehrmannManteuffelTancrediWeihs2014}. For further details of this approach we refer the reader to J.Henn contribution in these proceedings. We only note that devising the universal algorithm of finding such simplified form of the differential system is very desirable.

\section{Dimensional recurrence relations}
Dimensional recurrence relations (DRRs) introduced by Tarasov \cite{Tarasov1996} (see also an earlier work \cite{DerkachovHonkonenPismak1990} for the specific three-loop case) can be derived in many ways, see original work by Tarasov \cite{Tarasov1996} and also Ref. \cite{Lee2010a}. It is very easy to derive raising DRR from the representation \eqref{eq:FPG}. Using the first relation \eqref{eq:Iproperties} we obtain $\tilde{J}^{(d-2)}(\mathbf{n})=\frac{1}{d/2-1}(G(A)\tilde{J}^{(d)})(\mathbf{n})$. Now, using the IBP identity \eqref{eq:ibp} with $Q_\alpha=G z_\alpha,\,Q=z_\alpha\partial_\alpha G=(L+1)F+L U$, we obtain
\begin{equation}
J^{(d-2)}(\mathbf{n})=U(A_1,\ldots,A_N)J^{(d)}(\mathbf{n})\,.
\end{equation}
Note that this formula works also for the integrals with numerators. Using this formula for the master integrals and performing the IBP reduction of the right-hand side, we obtain the system of dimensional recurrences
\begin{equation}
\mathbf{J}^{(d-2)} =\mathbb{M}(\epsilon) \mathbf{J}^{(d)}\,.
\end{equation}
The general solution of this system depends on several arbitrary periodic functions $\omega^i(d)=\omega^i(d-2)$ (compare with arbitrary constants in the general solution of differential equations). In order to fix those functions, one may apply the DRA method \cite{Lee2010}. The basic idea of this method is the following. Each $\omega^i$, understood as the function of $z=\exp(i\pi d)$, appears to be a meromorphic function on $\overline{\mathbb{C}}$. Therefore, in order to fix it, it is sufficient to know the position of its poles and principal parts of the Lorent expansion around them (plus one constant). This information can be extracted from the consideration of the singularities of the master integrals on any vertical stripe with width 2 on the complex plane of $d$. For the details and examples of application of the DRA method we refer the reader to Refs. \cite{Lee2010,LeeTer2010,Lee2011e,LeeSmirnov2012}.

\section*{Acknowledgments}
This work has been supported in part by the Ministry of Education and Science of the Russian
Federation and RFBR grant No. 13-02-01023.

\section*{References}

\begin{thebibliography}{10}

\bibitem{Chetyrkin1981}
K.~G. Chetyrkin and F.~V. Tkachov.
\newblock {\em Nucl. Phys.}, B192:159, 1981.

\bibitem{Tkachov1981}
F.~V. Tkachov.
\newblock {\em Phys. Lett. B}, 100(1):65, 1981.

\bibitem{Laporta2000}
S.~Laporta.
\newblock {\em Int.~J.~Mod.~Phys.~A}, 15:5087, 2000.

\bibitem{Kotikov1991}
A.~V. Kotikov.
\newblock {\em Phys. {L}ett. B}, 254:158, 1991; 259:314, 1991; 267:123, 1991.

\bibitem{Remiddi1997}
E. Remiddi.
\newblock {\em Nuovo {C}im.}, A110:1435, 1997.

\bibitem{Tarasov1996}
O.~V. Tarasov.
\newblock {\em Phys.~Rev.~D}, 54:6479, 1996.

\bibitem{Lee2010}
R.N. Lee.
\newblock {\em Nucl. Phys. B}, 830:474, 2010.

\bibitem{DerkachovHonkonenPismak1990}
S.~E. Derkachov, J.~Honkonen, and Y.~M. Pis'mak.
\newblock {\em J. Phys. A}, 23:5563, 1990.

\bibitem{LeePomeransky2013}
R.~N. Lee and Andrei~A. Pomeransky.
\newblock {\em JHEP}, 1311:165, 2013.

\bibitem{Gehrmann2000}
T.~Gehrmann and E.~Remiddi.
\newblock {\em Nucl.~{P}hys.~{B}}, 580:485, 2000.

\bibitem{Henn2013}
J.~M. Henn.
\newblock {\em Phys.Rev.Lett.}, 110(25):251601, 2013.
\bibitem{HennSmirnov2013}
J.~M. Henn and V.~A. Smirnov.
\newblock {\em JHEP}, 1311:041, 2013.

\bibitem{HennSmirnovSmirnov2014}
J.~M. Henn, A.~V. Smirnov, and V.~A. Smirnov.
\newblock {\em JHEP}, 1307:128, 2013; 1403:088, 2014.

\bibitem{HennMelnikovSmirnov2014}
J.~M. Henn, K. Melnikov, and V.~A. Smirnov.
\newblock 2014.

\bibitem{CaolaHennMelnikovSmirnov2014}
F. Caola, J.~M. Henn, K. Melnikov, and V.~A. Smirnov.
\newblock {\em arXiv:}1404.5590,  2014.


\bibitem{GehrmannManteuffelTancrediWeihs2014}
T. Gehrmann, A. von Manteuffel, L. Tancredi, and E. Weihs.
\newblock 2014.

\bibitem{Lee2010a}
R.~N. Lee.
\newblock {Nucl. Phys. B} 205-206:135, 2010.



\bibitem{LeeTer2010}
R.~N. Lee and I. S. Terekhov.
\newblock {\em JHEP}, 1101:068, 2011.

\bibitem{Lee2011e}
R.~N. Lee, A.~V. Smirnov, and V.~A. Smirnov.
\newblock {\em JHEP}, 2010:1, 2010;
\newblock {\em Eur.Phys.J.}, C71:1708, 2011;
\newblock {\em Nucl.Phys. B}, 856:95, 2012.

\bibitem{LeeSmirnov2012}
R.~N. Lee and V.~A. Smirnov.
\newblock {\em JHEP}, 1102:102, 2011;
\newblock {\em JHEP}, 1212:104, 2012.

\end{thebibliography}

\end{document}